# Theoretical Prediction of Low-Energy Photoelectron Spectra of Al$_n$Ni$^-$ Clusters (n=1-13)

Paulo H. Acioli

†Department of Physics, Northeastern Illinois University, Chicago, IL 60625, USA

**ABSTRACT:**

**Context**

Mixed-metal clusters have long been studied because of their peculiar properties and how they change with cluster size, composition and charge state and their potential roles in catalysis. The characterization of these clusters is therefore a very important issue. One of the main experimental tools for characterizing their electronic structure is photoelectron spectroscopy. Theoretical computation completes the task by fully determining the structural properties and matching the theoretical predictions to the measured spectra. We present density functional theory computations of the structural, magnetic and electronic properties of negatively charged mixed Al$_n$Ni$^-$ clusters with up to 13 Al atoms. The lowest energy structures of the anionic clusters with up to 7 atoms are also found to be low energy isomers of the neutral counterparts found in the literature. The 13-atom cluster is found to be a quartet and a perfect icosahedron. The predicted photoelectron spectra are also presented and can be used to interpret future experimental data. We also presented indicators that can be used to determine the potential of these systems for single atom catalysis. These indicators point to smaller clusters to be more reactive as the gap between the Fermi energy and the center of the d-band increases with cluster size and that Ni occupy and internal site for n=11-13. We speculate that reactivity can be enhanced if one adds an additional Ni atom.

**Methods**

The DFT calculations were performed using the Becke exchange and Perdew-Wang/91 correlation functionals (BPW91), a DFT-optimized all-electron basis set for the aluminum atom and the Stuttgart small core pseudopotential for the Ni atom. All of the computations used the Gaussian 03 software.

**KEYWORDS:** Photoelectron spectroscopy, electron binding energies, size effects, structure effects, composition effects, DFT, single atom catalysis tuning.

# Introduction

The field of clusters is very mature but it still very important in helping us understand the change in properties from small molecules up to bulk size materials [1-3]. Clusters of transition metal element are of particular importance for its role in the field of catalysis [4,5], in particular single-atom catalysis [6,7]. Mixed-metal clusters are a fertile ground to design new materials where one can tune properties by changing the composition, charge state, and or the size of the cluster. In reference [8] we studied the roles of size, structure and composition in 13-atom clusters of pure aluminum and mixed Al-Ni clusters. More recently we studied how supporting Al clusters can change the nature of the frontier orbitals of Pt [9], an important element in the catalysis field. It is then important to continue to characterize mixed-metal clusters both experimentally and theoretically. One of the important experimental tools for characterization of clusters of different sizes is the use of photoelectron spectroscopy (PES) [10-12]. The synergy between PES and computational have been instrumental in determining the structural, electronic, as well as magnetic properties of clusters. In the past few years we have studied mixed-metal clusters using density functional theory computations of the photoelectron spectra in collaboration with experiment to determine the preferred structure and electronic properties of mixed $Al_nMo$ and $Al_nPt$ clusters [9,13-15].

In the present work we present a thorough exploration of the lowest energy structures of mixed aluminum-nickel clusters with up to 14 atoms. Computational studies of the electronic structure of these systems remain relatively limited [16-19] and are focused on Ni rich clusters. Other theoretical studies of mixed Al-Ni clusters use semi-empirical potentials [20-24] and were focused on the structural properties of these clusters. Here we will focus on the determining low-lying isomers, electronic structure, and magnetic properties of $Al_nNi^-$ clusters (n=1-13). More importantly we will present the photoelectron spectra of these clusters for comparison with future experiments in these systems when available.

**Computational and Methodological Aspects**

The calculations were performed within the gradient-corrected density functional theory (DFT). We used the Becke exchange [25] and Perdew-Wang/91 correlation [26] functionals (BPW91), a DFT-optimized all-electron basis set [27] for the aluminum atom and the Stuttgart small core pseudopotential for the Ni atom [28]. Unrestricted wave functions were utilized in all computations presented here. These choices were made thorough tests on several properties of the Ni, Al atoms and the dimers $Al_2$, $Ni^2$, and AlNi. The results of these tests are presented in Ref. [8]. All of the computations used the Gaussian 03 software [29].

The structural optimizations were carried out using gradient-based techniques with no symmetry constraints imposed. Normal mode analysis was applied to the stationary configurations obtained to separate stable equilibrium structures from transition state conformations. For the stable structures, we computed and analyzed the spectra of their electron binding energies. This was accomplished using a highly accurate scheme for conversion of the KS single-particle eigenenergies into true eigenenergies (or, alternatively, binding energies) of electrons. [30] The EBE corresponding to the *i*-th KS orbital with eigenenergy $\varepsilon(i)$ is obtained as a sum

$$EBE(i) = -\varepsilon(i) + \Delta(i), \qquad (1)$$

where $\Delta(i)$ is a correction term. The methodology for calculating the corrections $\Delta(i)$ is rigorous in that it uses only ground state quantities well-defined within the traditional (time-independent) DFT, it is robust in that it is applicable to any version of DFT, and it yields orbital-specific corrections $\Delta(i)$ to the KS eigenenergies.[30]

**Results and Discussion**.

**Structural properties**

In Fig.1 we display low energy stable structural forms of $Al_nNi^-$ (n=1,13) found in this work. We have found different structural motifs depending on cluster size. We found a pyramid or capped pyramid for n=4-5. In the case of n=6-7, we have found either a bi-capped trigonal prism or distorted capped octahedron as the two competing motifs, with the Ni atom in different positions.

The Al$_6$Ni$^-$ structure is similar to the result obtained for Al$_6$Pt$^-$[13] but very different from what we found for Al$_6$Mo$^-$[14], where the lowest structure was a distorted octahedron of D$_{3d}$ symmetry. This is not surprising as the neutral counterpart, Al$_6$Mo, has a closed shell and can be seen as a magic number cluster with perfect O$_h$ symmetry[14] and Ni and Pt atoms have similar valence electronic structure, with [Kr]4d$^5$5s$^1$ and [Xe]4f$^{14}$5d$^9$6s$^1$ configurations, respectively. Some of the isomers for Al$_n$Ni- for n=1-7 can be found as a possible isomer of the neutral counterpart found in Ref. [31]. In the case of n=8-9 we found capped octahedron or fused octahedra as well as the precursors of the icosahedron that is the lowest energy isomer of Al$_{12}$Ni$^-$. In all of the sizes up to 9 Al atoms the Ni atom lies on a surface site. We searched for packed structures with the Ni atom occupying an internal position, but these were high energy isomers that were not included in this work. The remainder cases are either precursors of the icosahedron or a derivative of the icosahedron such as Al$_{13}$Ni$^-$. Al$_{11}$Ni$^-$ was the first case where the most stable isomer had a Ni at an interior site.

The binding energy per atom of the clusters defined as

$$BE_n(\text{Al}_n\text{Ni}^-) = \left[ nE(\text{Al}_n) + E(\text{Ni}^-) - E(\text{Al}_n\text{Ni}^-) \right] / n \qquad (2)$$

are presented in Fig. 2. The binding energy increases non-monotonically as function of size starting from 1.218 eV/atom for AlNi$^-$ and ending with 2.757 eV/atom for Al$_{13}$Ni$^-$. These are lower than the measured values 3.39 eV/atom and 4.44 eV/atom of bulk aluminum and nickel, respectively. One can see that n=5 and n=11 seem to be slightly less stable as its neighbors. To better visualize the relative stability of these clusters as a function of size we also computed the second energy difference as

$$\Delta E_2(\text{Al}_n\text{Ni}^-) = E(\text{Al}_{n+1}\text{Ni}^-) + E(\text{Al}_{n-1}\text{Ni}^-) - 2E(\text{Al}_n\text{Ni}^-). \qquad (3)$$

Fig. 3 does confirm that the Al$_5$Ni$^-$ and Al$_{11}$Ni$^-$ are less stable than its neighbors. We also note that Al$_8$Ni$^-$ is a less stable than its neighbors. We can understand this by the fact that for n=6, 9, and 12, there is a change in the structural motif of the more stable isomer. Al$_{12}$Ni$^-$, in particular is a perfect icosahedron that is expected to be more stable than its precursor.

Next, we present the magnetic moments of the lowest energy isomers of $Al_nNi^-$ (n=1,13) in Fig. 3. It is interesting to note that in most cases the magnetic moments oscillate between 0 and 1μB. The only exceptions are n=5 and n=12 where the magnetic moments are 2μB and 3μB, respectively. The latter agrees with the results we presented in our earlier work [8] in which we used $Al_{12}^-$, $Al_{11}Ni^-$, $Al_{12}Ni^-$, and $Al_{13}^-$ as an example to interpret the photoelectron spectra of pure and mixed clusters, disentangling the roles of charge state, spin state, size, and composition. In Ref. [31] The authors present the magnetic moments for neutral $Al_nNi$ (n=1,7) clusters, they also observe an alternation between 0 and 1μB, starting with 1μB for AlNi. The only exception of this alternation in Ref. [31] is the case of $Al_4Ni$ that has a moment of 2μB.

**Electronic Properties**

Photoelectron spectra is a measurement of the binding energy of the electrons, in this work we compute the zero temperature electron binding energies of mixed $Al_nNi^-$ (n=1,13). In Figs. 4 we present the results for the lowest energy isomers of these systems. In the Supplemental Material we present the results for all isomers presented in Fig. 1. The lowest bound electron in the dimer ($AlNi^-$) has an electron binding energy of 1.165 eV. As a comparison the electron affinities of Al, Ni, $Al_2$ and $Ni_2$ are 0.441 eV, 1.156eV, 1.46eV, and 0.926, respectively. In Table 1 we present the percentage of S, P and D character in the HOMO of the lowest energy structures of $Al_nNi^-$. A close inspection of the highest occupied molecular orbital (HOMO) of $AlNi^-$, shows that this orbital is dominated by one of the Ni 5d orbitals, thus justifying the closeness of the values of the EBE of $AlNi^-$ and the electron affinity of Ni. The value of the lowest bound electron binding energy increases with size up to $Al_6Ni^-$ when it reaches the value of 2.474eV eV. With the change in motif of $Al_7Ni^-$, we observe a small reduction on the first detachment energy. The first detachment energy continues to increase with cluster size up to $Al_{13}Ni^-$. When we analyze the whole spectrum in the energy range considered in this work (up to 3.4eV) the spectra for the smaller sizes seem to indicate prominent peaks for the first and second detachment energies for n=1, 3, and 7. For the other sizes there are many lines that would seem to create broad peaks. This effect will be further enhanced by the inclusion of the other low-energy isomers as discussed in the Supplemental Material. These results can be used to interpret future PES experiments. For a better understanding on how these results can be used refer to Ref. [8].

The electronic structure of these clusters can also help us understand the possible role in single-atom catalysis as pointed out in Ref. [9]. For instance, one can think that the aluminum clusters can be used to tune the catalytic properties of Ni. This can be understood by the change in the electronic properties through many different indicators. The first one is the vertical electron detachment of the lowest bound electron (1st VDE) that is shown in Fig. 6. As the cluster grows in size the 1st VDE increases nonmonotonically with a local maximum at n=7. Another interesting indicator is the charge on the Ni atom (see Fig. 7). The charge varies slightly in the range n=1-10, varying between -0.57 and -0.92$e$. One notices a sharp drop at n=11, and a further decrease for n=12,13, to a maximum charge of -2.75$e$. This is a clear indication of the change in structure where the Ni atom goes to an interior site on the icosahedron motif. One should not expect Ni to play a role in catalysis in these sizes. This abrupt change in charge in the Ni atom is accompanied by an equivalent change in charge of the outside shell that now has a net positive charge.

The next set of tunable properties involve the frontier orbital energetics and the d-character of the valence orbitals as discussed in Ref. [9]. Using the same population analysis that allowed us to determine the d-character of the HOMO shown on Table 1 we determine the s-, p-, and d-character of the valence orbitals. Because, catalytic properties are usually associate with the d-orbitals of the transition metal, here we will only present an analysis of the d-character. Like in Ref. [9] we defined the Fermi energy as an average of the HOMO and LUMO (lowest unoccupied molecular orbital). We also define the center of the d-band as a weighted average of the occupied valence orbitals using as weights the percentage d-character. In Fig. 8 we present the Fermi energy and the position of the center of the d-band as function of cluster sizes. We can see that there is a non-monotonic decrease in both cases, which is similar to the trend of the 1st VDE shown in Figs.

5 and 6. Under Koopman's theorem the 1$^{st}$ VDE is associated with the negative of the energy of the HOMO. A better indicator of reactivity is the gap between the frontier orbitals (HOMO-LUMO) or the gap between the Fermi energy and the center of the d-band. These are presented in Fig. 9. The HOMO-LUMO gap shows an alternation ranging between 0.23 and 0.66 eV for n=1-12, with a sharp increase for Al$_{13}$Ni$^-$ where the value is 1.60 eV. This can be attributed to the closed packed nature of the Al$_{12}$Ni$^-$ structure that is a perfect icosahedron. The picture of the gap between the Fermi energy and the center of the d-band is slightly different. The gap increases rapidly from n=1 to n=3, and then shows some alternance up to n=10, when it starts to increase for a maximum of about 2.36eV at n=13. Although the numerical values are different for both gaps the picture seems to be similar that the smaller sizes should be more reactive, following the pattern that for the larger sizes Ni occupies an interior site in the cluster.

**Conclusions**

We present structural, magnetic and electronic properties of Al$_n$Ni$^-$ clusters with up to 14 atoms. The structures found in this work are similar to previous structures found for neutral atoms using DFT [30] for n=1,7. The 13 atom cluster (Al$_{12}$Ni$^-$) has been studied previously using DFT and semi-empirical methods and they all are in agreement with our lowest energy perfect icosahedron as well as our prior study [8] of using the 13 atom cluster as a model system to understand the effects of composition, size, charge state in explaining particular trends in the photoelectron spectra of these systems. Some of the small size structures (n<7) were also found as stable isomers of Al$_n$Mo$^-$[15] and Al$_n$Pt$^-$[13]. One notable exception was Al$_6$Mo$^-$ [14], where the most stable structure was a distorted octahedron with Mo occupying the central site and this structure was not found to be a low-lying structure in either Al$_6$Ni$^-$ or Al$_6$Pt$^-$.

We also present the electron binding energy that can be used to corroborate our findings when experimental photoelectron spectra are made available. For the small sizes the 1$^{st}$ VDE starts at around 1.2 eV and grows non-monotonically with cluster sizes. The second VDE shows different gaps as a function of cluster size showing the largest gap for Al$_3$Ni$^-$. The data supplied in Fig. 5 can be used to interpret future experimental results. It is worth noting that our results are in very good agreement with prior theoretical and experimental results for $Al_{11}Ni^-$ and $Al_{12}Ni^-$ [8].

Finally, we also present results of how one can use the $Al_nNi^-$ clusters to tune in different indicators of catalytic behavior and we conclude that the smaller sizes should be more reactive, while the close packed structures (n=11-13) should less reactive due to the center of the d-band sinking well below the frontier orbitals and the Ni atom occupying a central position on the cluster. One can speculate that mixed clusters with more than one Ni atom might be more appropriate for catalysis even for larger sizes, as the second Ni should occupy a surface site.

## ASSOCIATED CONTENT

**Supporting Material**

Results of the photoelectron spectra of higher energy isomers of $Al_nNi^-$.

## AUTHOR INFORMATION

**Corresponding Author**

* E-mail: p-acioli@neiu.edu (Paulo H. Acioli)

**Notes**

The authors declare no competing financial interest.

## ACKNOWLEDGMENTS
This work was supported by the Northeastern Illinois University Department of Physics and Astronomy.

**Funding Declaration**
No funding.

Table 1 – Percentage of S, P, and D character of the Highest Occupied Molecular Orbital of $Al_nNi^-$ (n=1,13).

| System | S(%) | P(%) | D(%) |
|---|---|---|---|
| $AlNi^-$ | 0 | 0 | 100 |
| $Al_2Ni^-$ | 2.7 | 66.3 | 31.1 |
| $Al_3Ni^-$ | 19.3 | 47.2 | 33.4 |
| $Al_4Ni^-$ | 11.4 | 4.7 | 83.9 |
| $Al_5Ni^-$ | 14.4 | 82.6 | 3.0 |
| $Al_6Ni^-$ | 4.7 | 58.2 | 37.2 |
| $Al_7Ni^-$ | 22.8 | 53.3 | 23.8 |
| $Al_8Ni^-$ | 19.1 | 67.1 | 13.8 |
| $Al_9Ni^-$ | 15.7 | 43.7 | 40.6 |
| $Al_{10}Ni^-$ | 33.9 | 60.4 | 5.7 |
| $Al_{11}Ni^-$ | 25.7 | 70.6 | 3.7 |
| $Al_{12}Ni^-$ | 4.8 | 94.9 | 0.3 |
| $Al_{13}Ni^-$ | 18.5 | 80.1 | 1.3 |

**Figure captions**

**Figure 1.** Stable isomers of $Al_nNi^-$ (n=1-13). For each isomer we present the binding energy per atom (eV/atom), its magnetic moment in µB, and its point group symmetry. The green atom represents Ni and the grey atoms represent Al.

**Figure 2.** Binding energy per atom (eV/atom) of the lowest energy isomer of $Al_nNi^-$ (n=1,13).

**Figure 3.** Second energy difference (eV) of the lowest energy isomer of $Al_nNi^-$ (n=1-13).

**Figure 4.** Total magnetic moment (µB) of the lowest energy isomer of $Al_nNi^-$ (n=1-13).

**Figure 5.** Computed electron binding energies lowest energy isomer of $Al_nNi^-$ (n=1-13).

**Figure 6.** First vertical detachment energy of the lowest energy isomer of $Al_nNi^-$ (n=1-13).

**Figure 7.** Charge on the Ni atom of the lowest energy isomer of $Al_nNi^-$ (n=1-13).

**Figure 8.** Fermi energy and the position of the center of the d-band of the lowest energy isomer of $Al_nNi^-$ (n=1-13).

**Figure 9.** HOMO-LUMO (blue line) and Fermi energy and the center of the d-band (red) gaps of the lowest energy isomer of $Al_nNi^-$ (n=1-13).

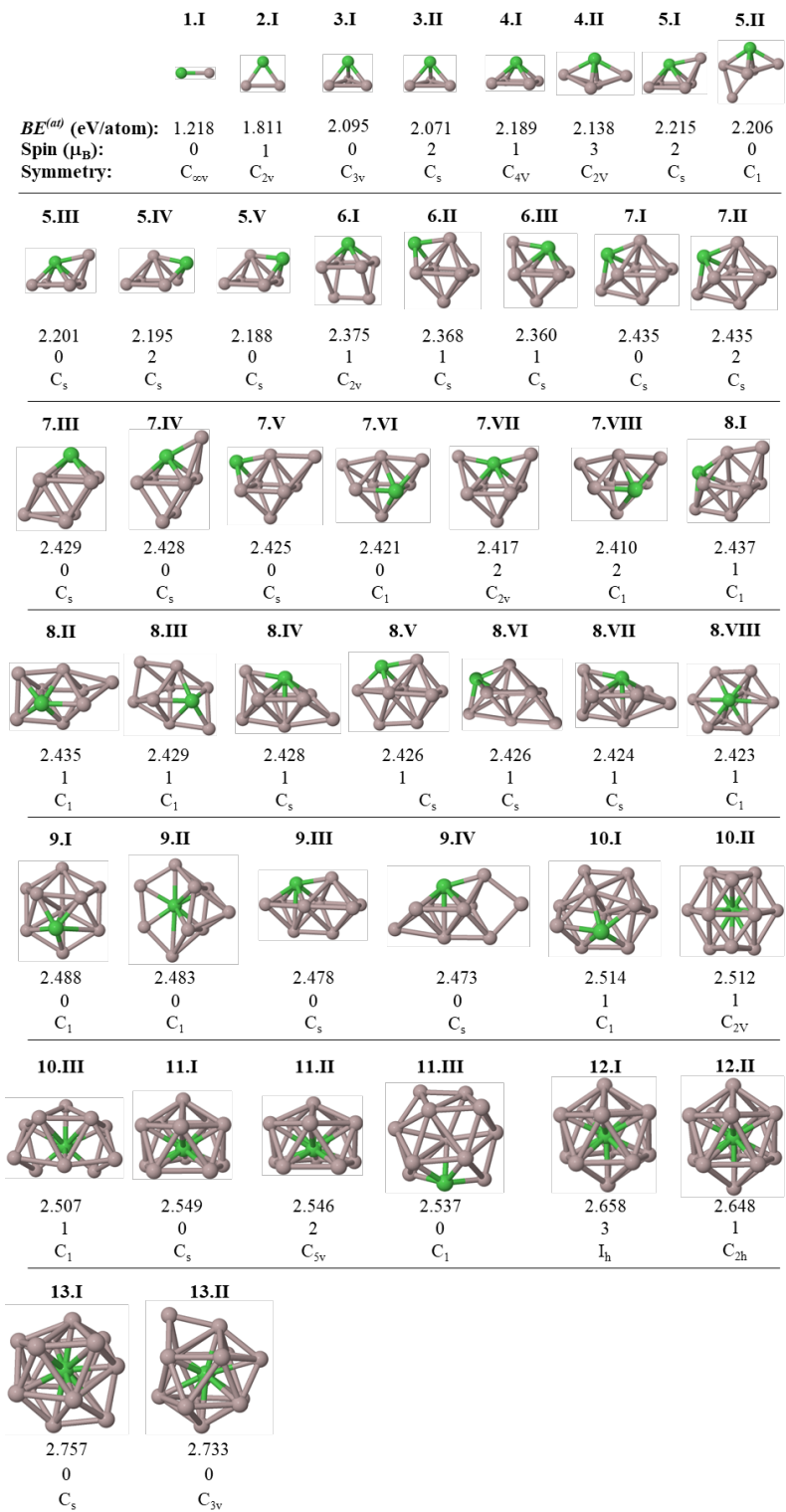

Fig.1

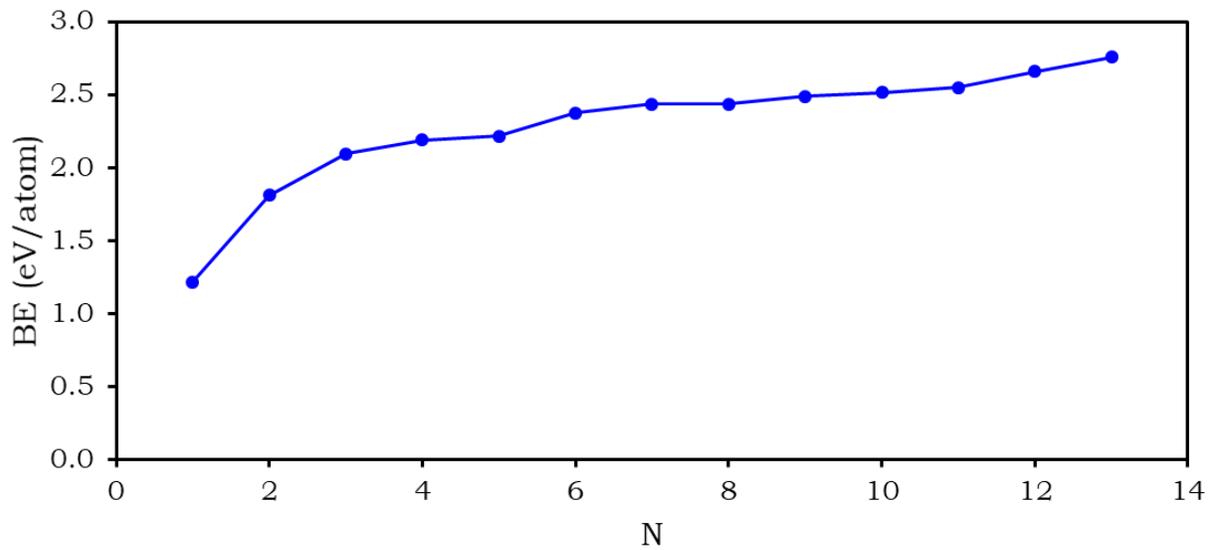

Fig. 2

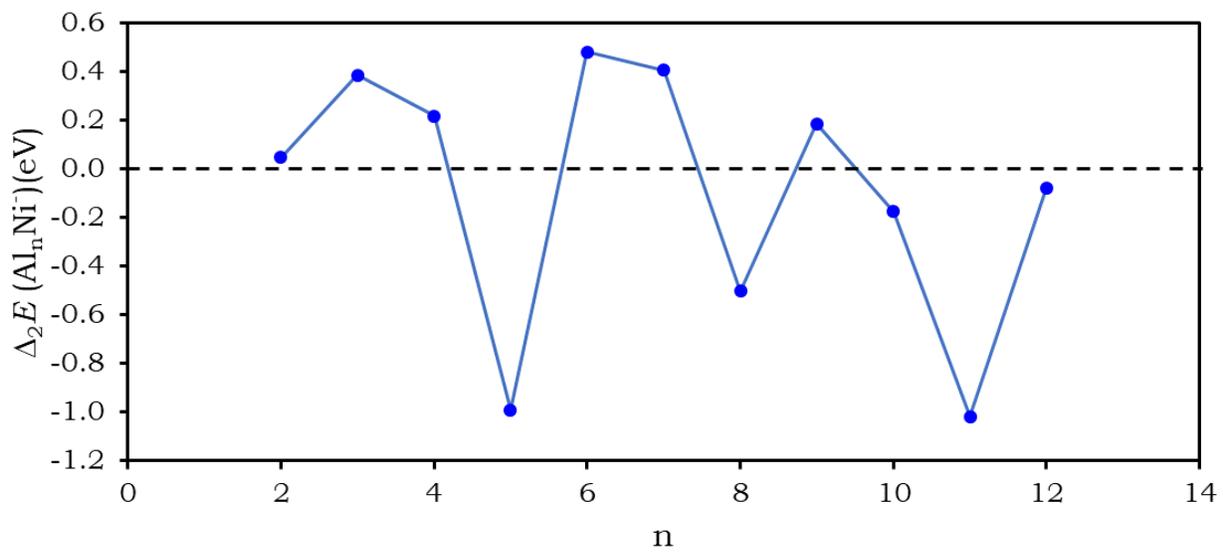

Fig. 3

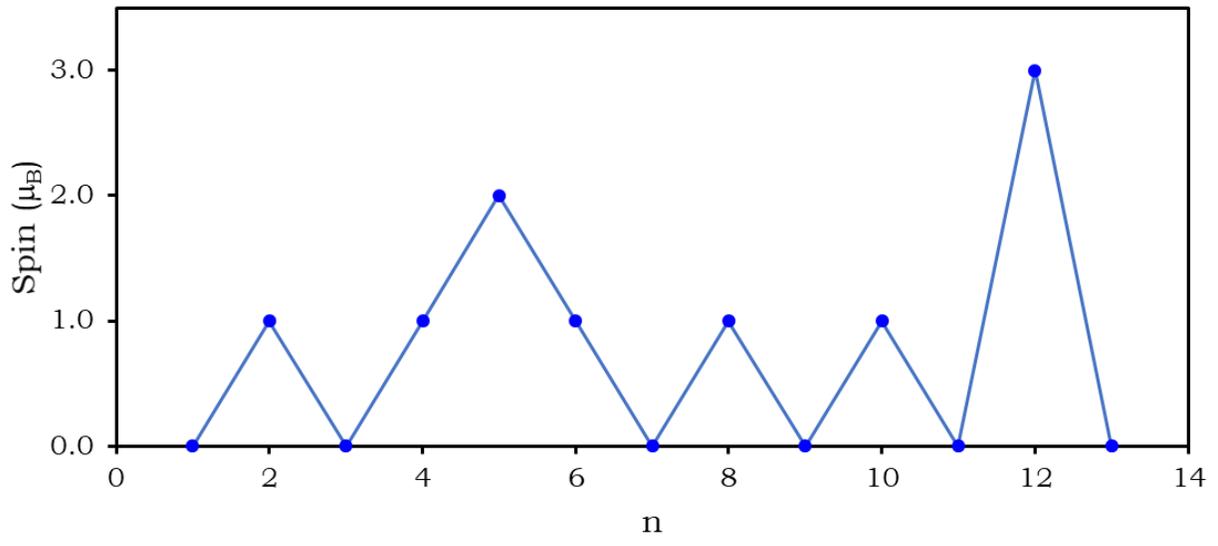

Fig. 4

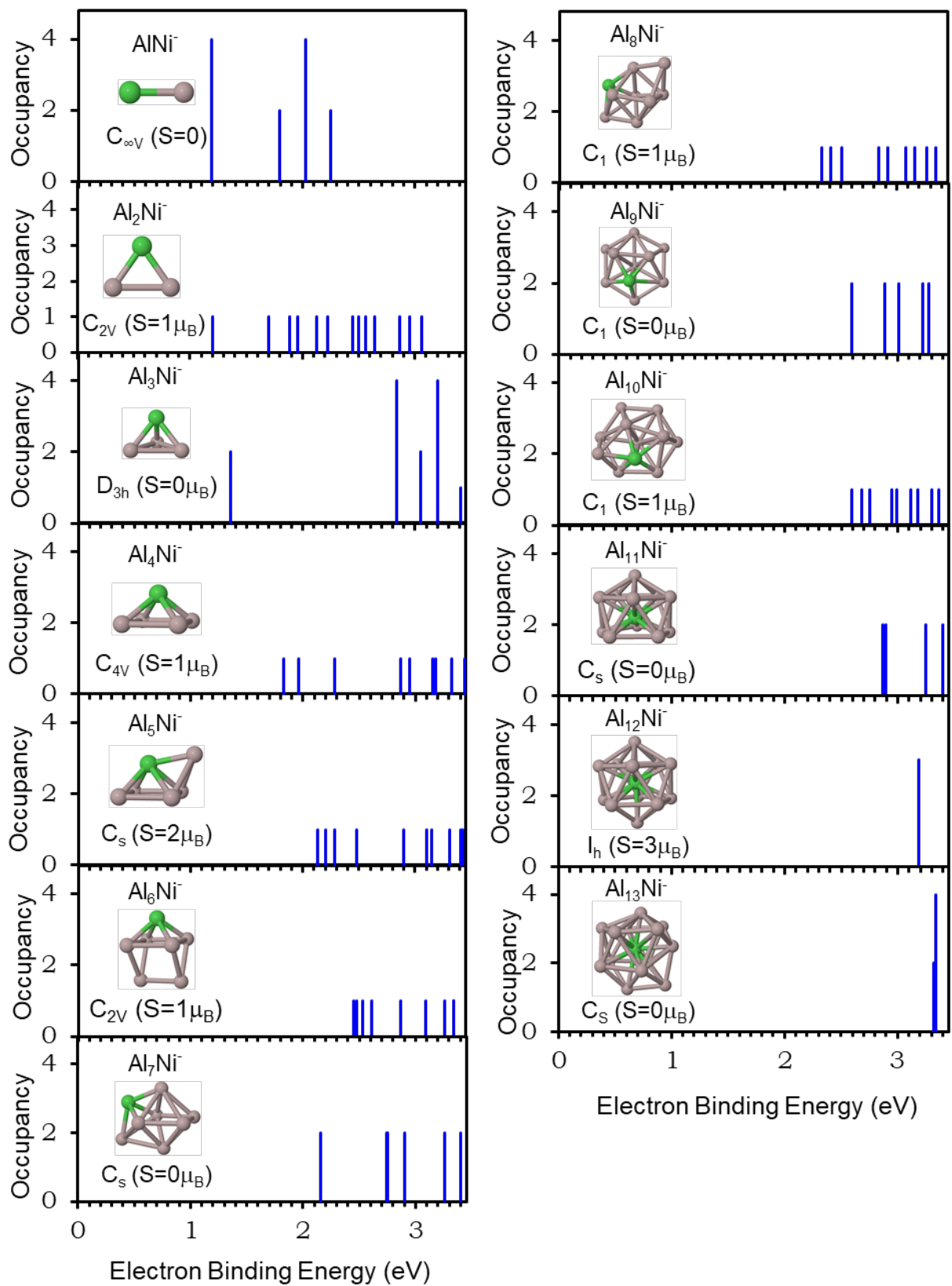

Fig. 5

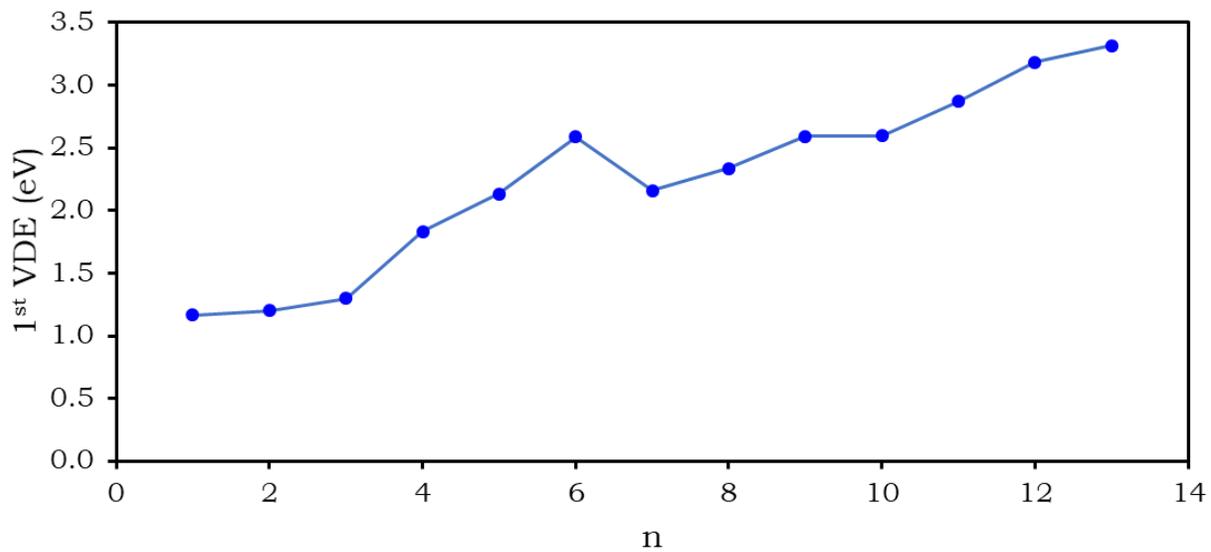

Fig. 6

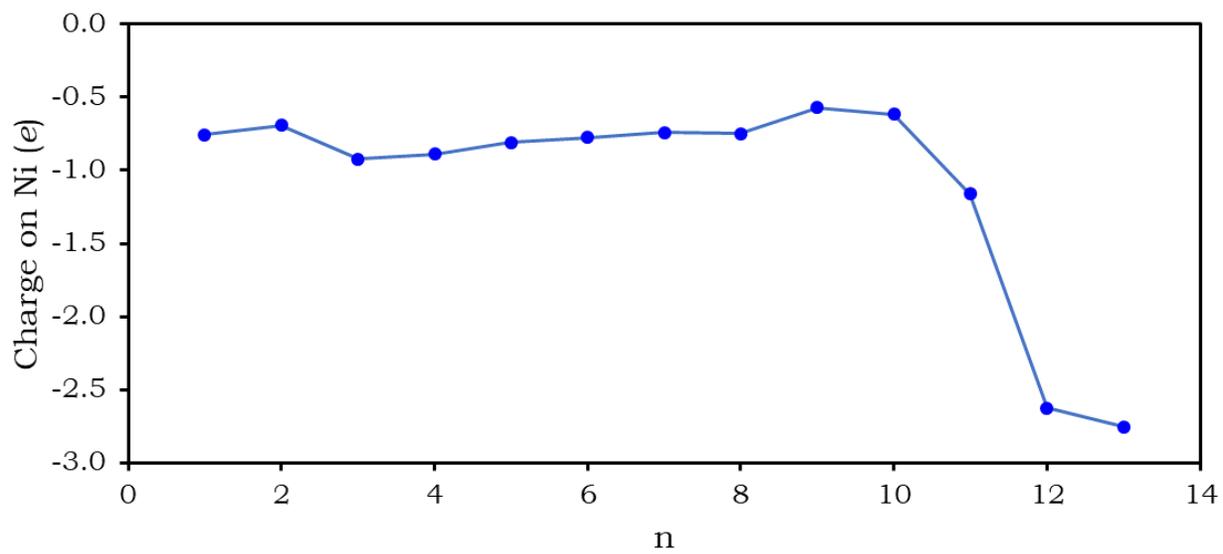

Fig. 7

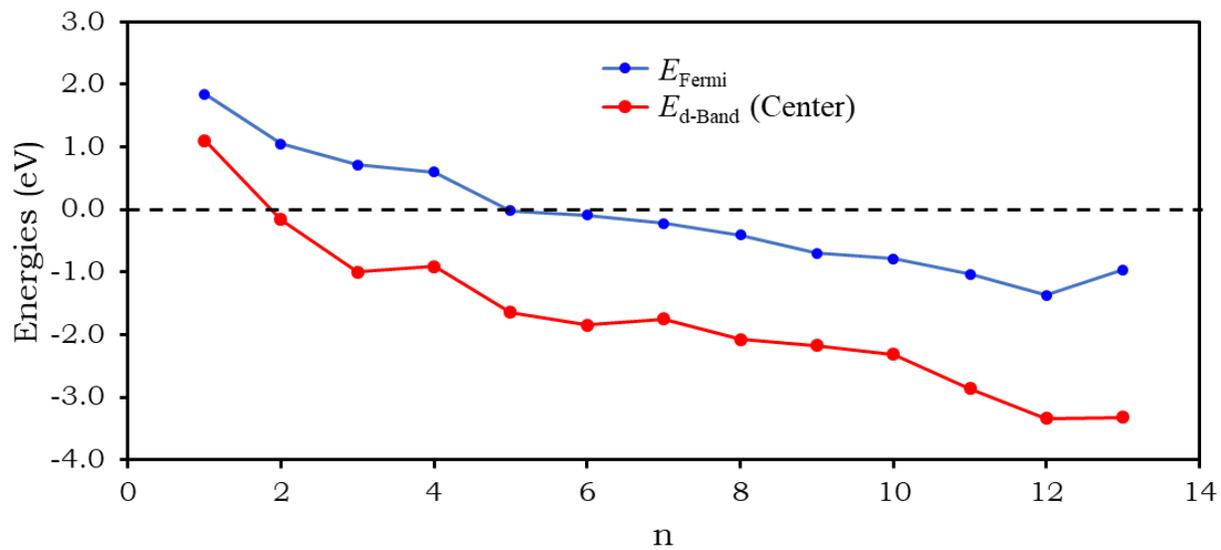

Fig. 8

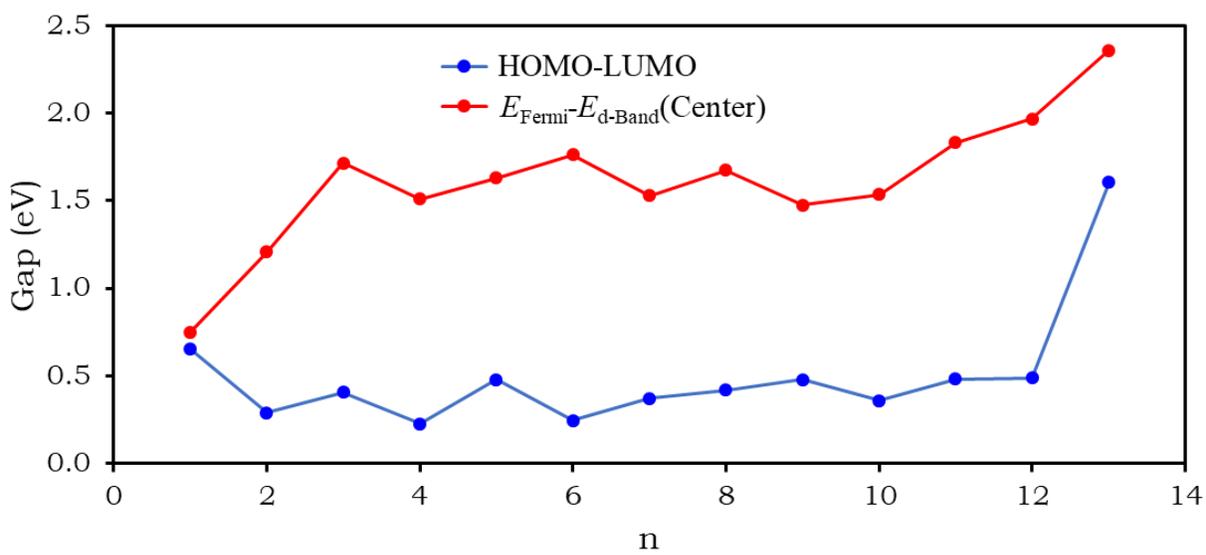

Fig. 9

# Supplemental Material
# Theoretical Prediction of Low-Energy Experimental Photoelectron Spectra of $Al_nNi^-$ Clusters (n=1-13)

Paulo H. Acioli
[†]Department of Physics, Northeastern Illinois University, Chicago, IL 60625, USA

In this supplemental material we present all electron binding energies for all the low-lying isomers presented in Fig. 1 of the main text.

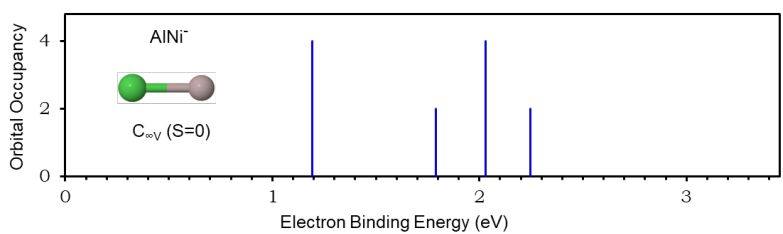

Fig. S1 – Electron binding energy of the lowest energy isomer of $AlNi^-$.

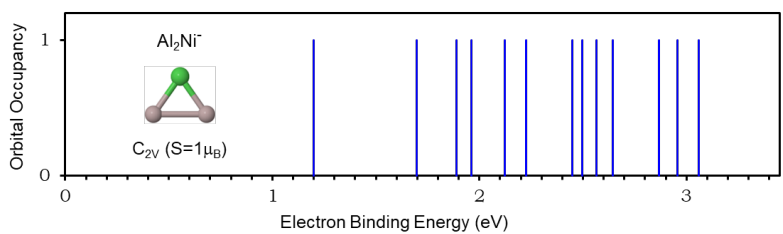

Fig. S2 - Electron binding energy of the lowest energy isomer of $Al_2Ni^-$.

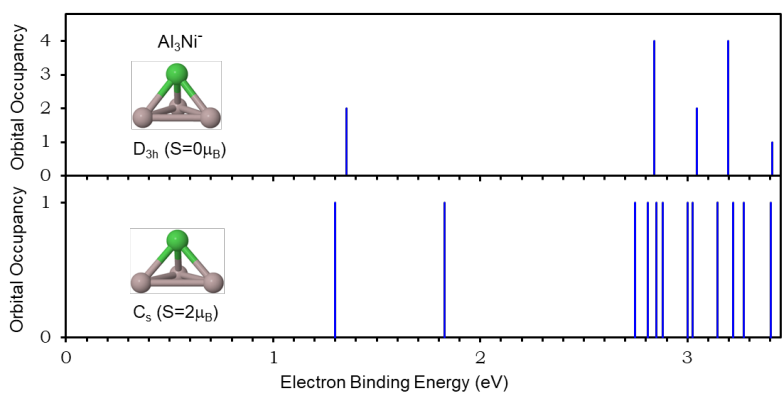

Fig. S3 - Electron binding energy of low energy isomers of Al$_3$Ni$^-$.

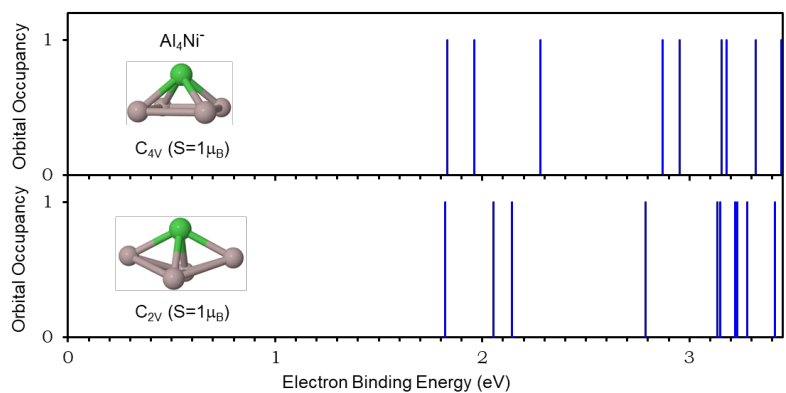

Fig. S4 - Electron binding energy of low energy isomers of Al$_4$Ni$^-$.

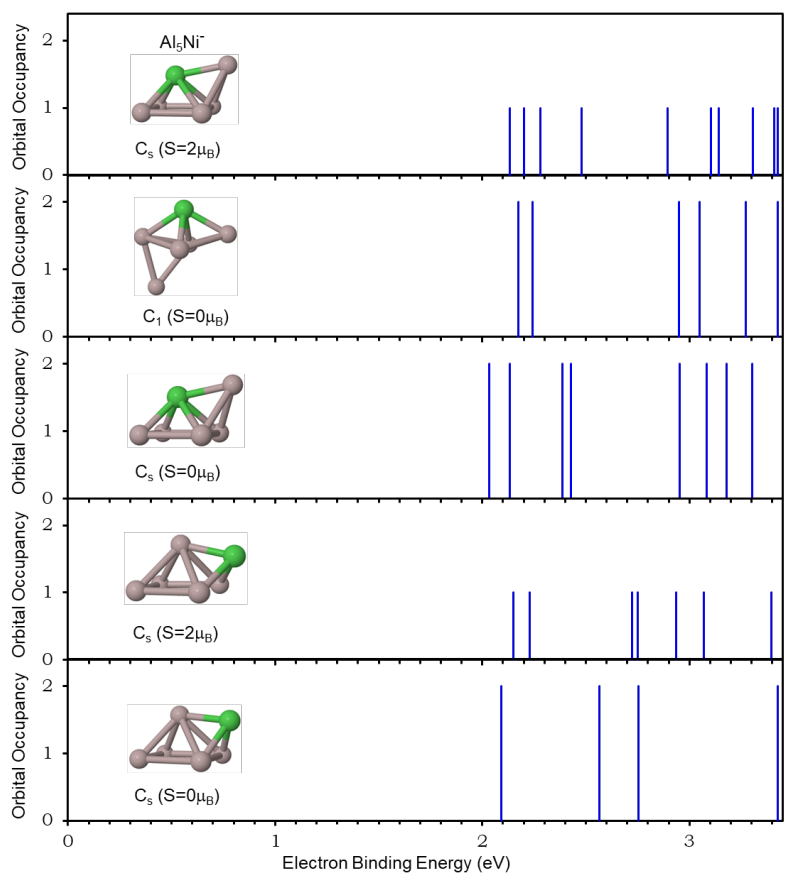

Fig. S5 - Electron binding energy of low energy isomers of Al$_5$Ni$^-$.

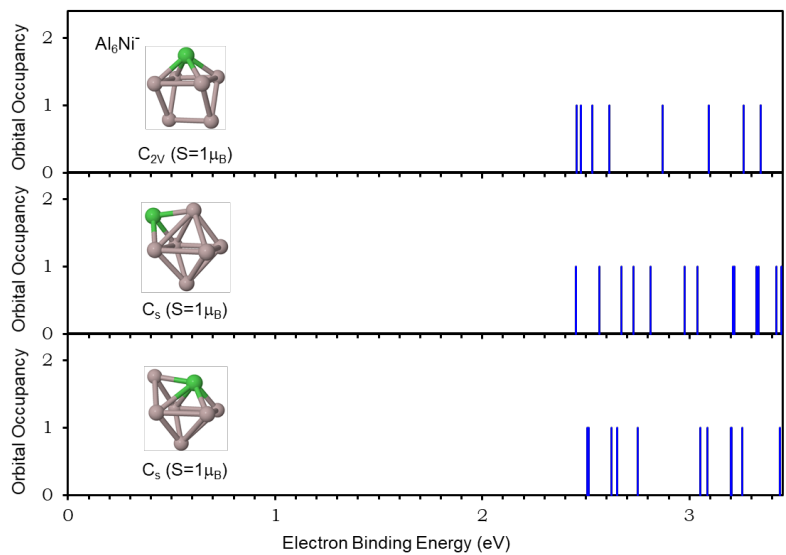

Fig. S6 - Electron binding energy of low energy isomers of Al$_6$Ni$^-$.

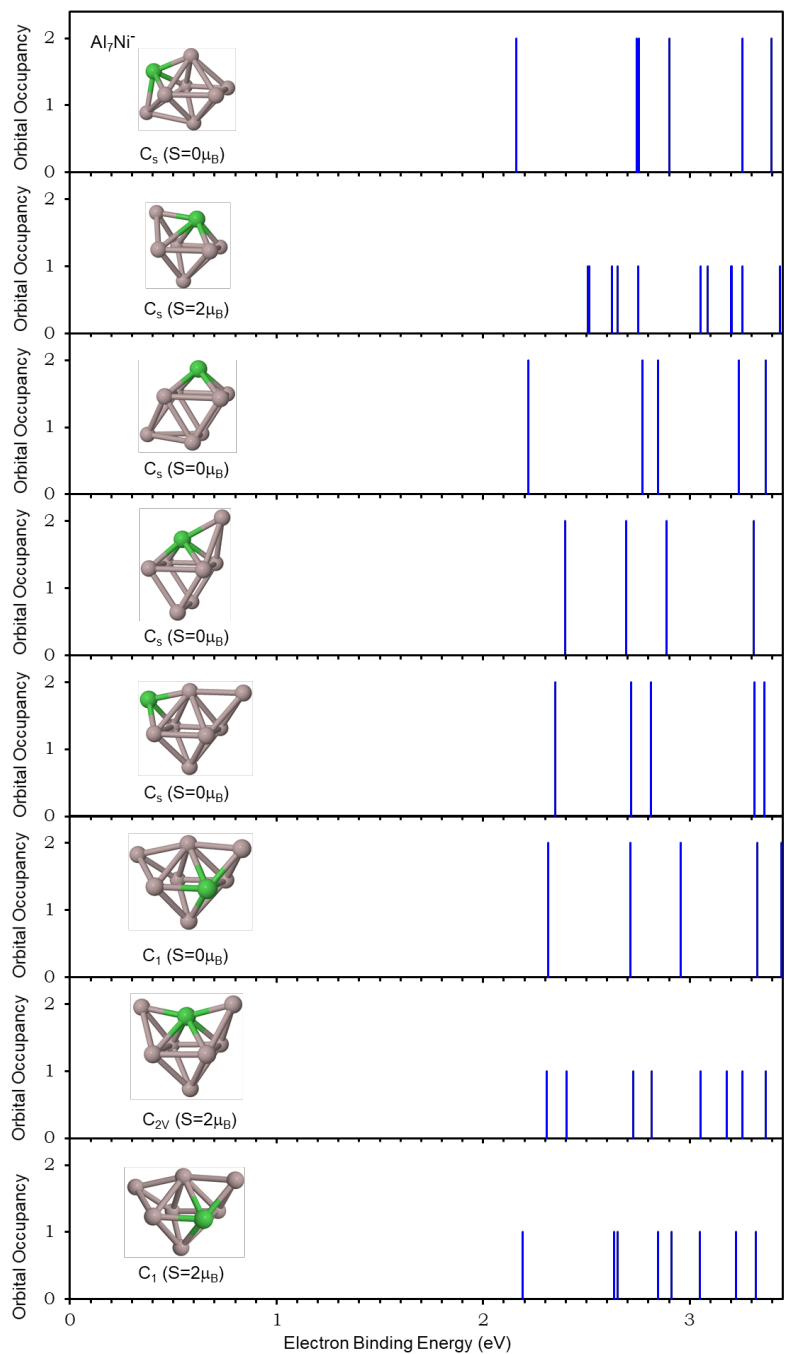

Fig. S7 - Electron binding energy of low energy isomers of $Al_7Ni^-$.

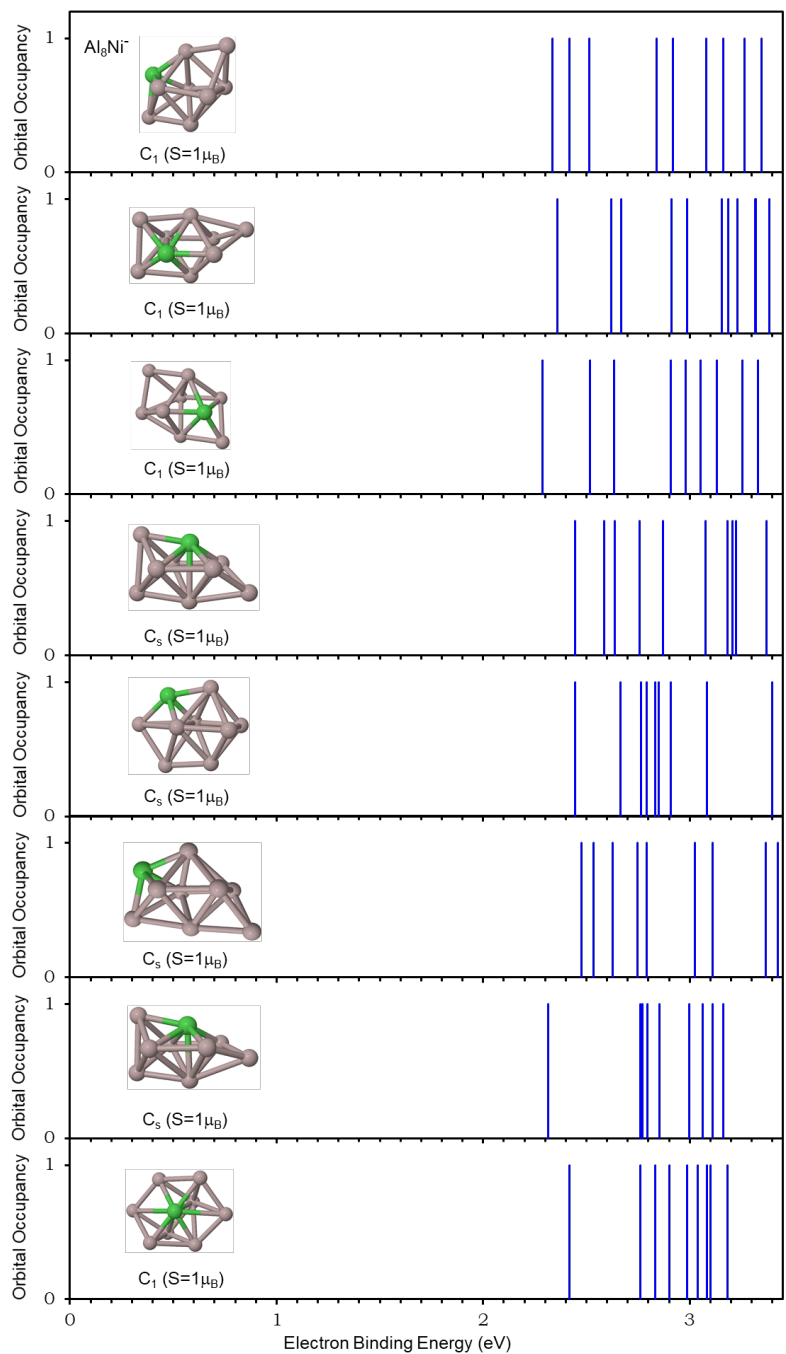

Fig. S8 - Electron binding energy of low energy isomers of $Al_8Ni^-$.

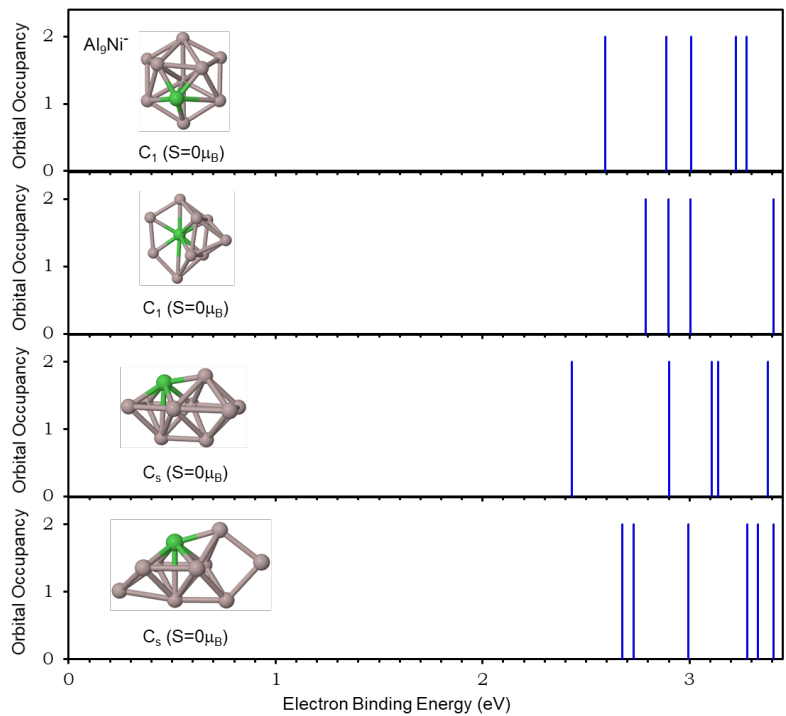

Fig. S9 - Electron binding energy of low energy isomers of Al$_9$Ni$^-$.

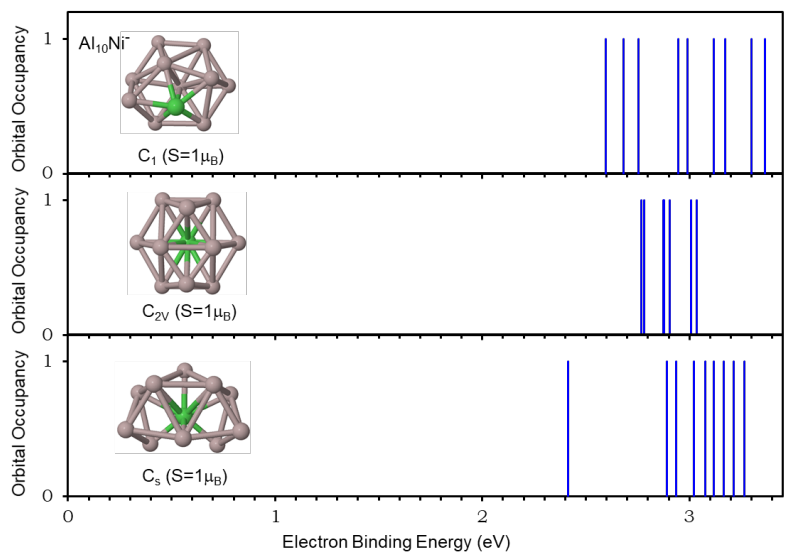

Fig. S10 - Electron binding energy of low energy isomers of Al$_{10}$Ni$^-$.

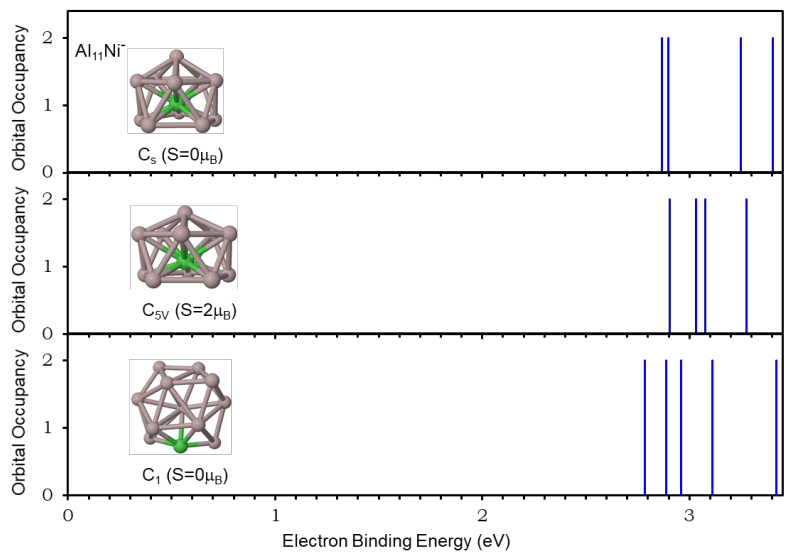

Fig. S11 - Electron binding energy of low energy isomers of $Al_{11}Ni^-$.

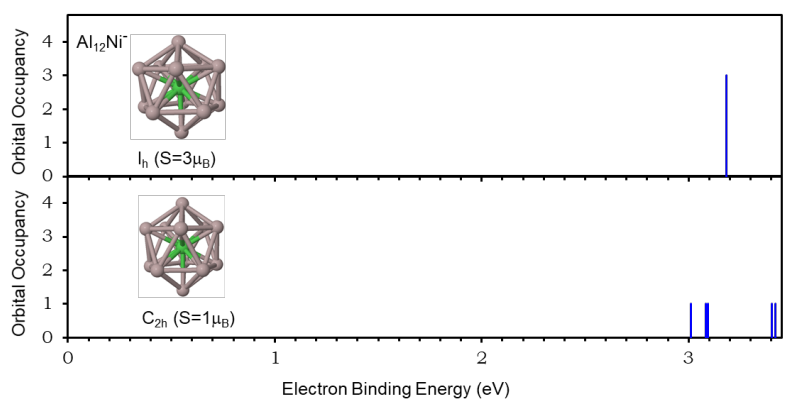

Fig. S12 - Electron binding energy of low energy isomers of $Al_{12}Ni^-$.

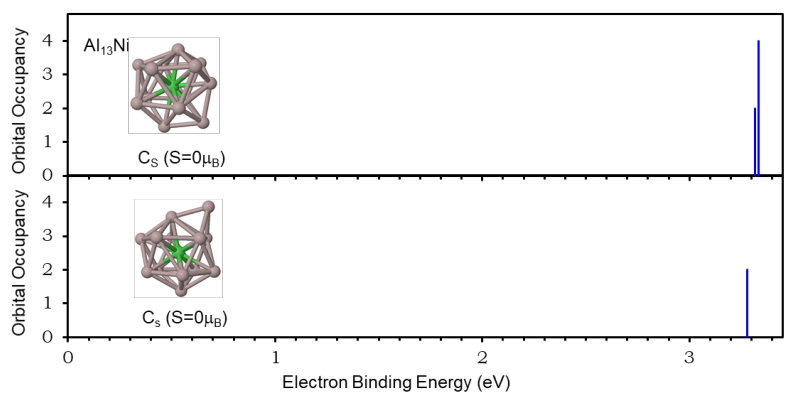

Fig. S13 - Electron binding energy of low energy isomers of $Al_{13}Ni^-$.